\documentclass{appolb}
\usepackage{graphicx}
\usepackage{hyperref}
\usepackage{todonotes}
\usepackage{amssymb}
\usepackage{amsmath}

\newcommand{\ex}[1]{\ensuremath{\times10^{#1}}}
\newcommand{\dmij}[1]{\ensuremath{\Delta m_{#1}^2}}
\newcommand{\Uaj}[1]{\ensuremath{|U_{#1}|^2}}
\newcommand{\nua}[1]{\ensuremath{\rlap{\kern-2.5pt\ensuremath{\overset{\scriptscriptstyle(-)}{\phantom{\nu}}}}{\ensuremath{{\nu}_{#1}}}}}
\newcommand{\Neff}{\ensuremath{N_{\rm eff}}}

\begin{document}

\title{Light sterile neutrinos: oscillations and cosmology%
\thanks{Presented at ``Matter To The Deepest'', XLIII International Conference of Theoretical Physics
--
Katowice/Chorz\'ow, Poland, 1--6 September 2019}%
}

\author{Stefano Gariazzo
\address{Instituto de F\'{\i}sica Corpuscular
(CSIC-Universitat de Val\`{e}ncia)\\
Parc Cient\'{\i}fic UV, C/ Catedr\'atico Jos\'e Beltr\'an, 2, E-46980 Paterna (Valencia), Spain}
}
\maketitle

\begin{abstract}
Light sterile neutrinos with a mass around 1 eV have been studied for many years as a possible explanation of the so called short-baseline neutrino oscillation anomalies. Recently, several neutrino oscillation experiments reported preferences for non-zero values of the mixing angles and squared mass differences for active-sterile mixing, which however are not always in agreement.
I review our current knowledge of the light sterile neutrino in the 3+1 model,
starting with a separate discussion on the status of the most relevant searches
and then analyzing the problems that arise when combining different probes in a global fit.
A short summary on the tension with cosmological observations is also provided.
\end{abstract}

\section{Introduction}
Our current knowledge of the oscillation parameters
in the three neutrino scheme has improved noticeably in the last twenty years,
see e.g.\ \cite{deSalas:2017kay}.
Yet, several anomalous experimental results remain unexplained.
Among these, we find the measurements by LSND \cite{Aguilar:2001ty},
plus the Gallium \cite{Giunti:2010zu,Kostensalo:2019vmv} and
reactor anomalies \cite{Mention:2011rk,Mueller:2011nm,Huber:2011wv}.
These anomalies, better discussed in the following sections,
might have a common explanation if a new neutrino eigenstate exists.
Oscillations between the three standard and the fourth neutrino,
driven by a new squared mass difference between the first and the fourth
neutrino mass eigenstates $\dmij{41}=m_4^2-m_1^2\simeq 1$~eV$^2$,
were proposed in order to give a common explanation of these anomalies.
The new neutrino, which cannot have standard model interactions \cite{ALEPH:2005ab},
is denoted as sterile.

The range at which oscillations between active and sterile neutrinos are dominant
is defined by the relation
\begin{equation}
\frac{\dmij{41} L}{E}\simeq1\,,
\end{equation}
where the quantity on the left hand side enters the neutrino oscillation formulas (see below).
For the distances $L$ and neutrino energies $E$ that match the above relation,
defining what are called Short BaseLine (SBL) oscillations,
the terms due to the solar and atmospheric mass splittings cannot develop,
and only the effect of \dmij{41} must be considered.
At SBL, therefore, one can write the
transition probability between a neutrino or antineutrino
of flavor $\alpha$ and one of flavor $\beta$
in the following way (see e.g.~\cite{Gariazzo:2015rra}):
\begin{eqnarray}
P^{\mbox{SBL}}_{\nua{\alpha}\rightarrow\nua{\beta}}
&\simeq&
\sin^2 2\vartheta_{\alpha\beta}\sin^2\left(\frac{\dmij{41}L}{4E}\right)
\,,
\qquad(\alpha\neq\beta)
\\
P^{\mbox{SBL}}_{\nua{\alpha}\rightarrow\nua{\alpha}}
&\simeq&
1-
\sin^2 2\vartheta_{\alpha\alpha}\sin^2\left(\frac{\dmij{41}L}{4E}\right)
\,,
\end{eqnarray}
where the effective angles $\vartheta_{\alpha\alpha}$ and $\vartheta_{\alpha\beta}$
depend on some elements of the fourth column of the (four by four) neutrino mixing matrix:
\begin{eqnarray}
\sin^22\vartheta_{\alpha\beta}
&=&
4\,\Uaj{\alpha4}\,\Uaj{\beta4}
\,,
\qquad(\alpha\neq\beta)
\\
\sin^22\vartheta_{\alpha\alpha}
&=&
4\,\Uaj{\alpha4}\, (1- \Uaj{\alpha4})
\,.
\end{eqnarray}
Since we generally expect the mixing matrix elements \Uaj{\alpha4} ($\alpha=e,\,\mu,\,\tau$)
to be small,
in order not to alter excessively the phenomenology of three-neutrino oscillations
observed in non-SBL experiments,
we expect the \emph{appearance} effective mixing angles $\vartheta_{\alpha\beta}$ ($\alpha\neq\beta$)
to be quadratically suppressed
with respect to the \emph{disappearance} ones, $\vartheta_{\alpha\alpha}$.
As we will discuss in the following, this is the reason for the existence
of the so-called appearance--disappearance tension.

\section{Disappearance constraints}
Let us now discuss more in details the various classes of experiments,
starting from disappearance probes.

\subsection{Electron (anti)neutrino disappearance}
The first anomaly in the electron neutrino disappearance channel
has been reported by the GALLEX and SAGE experiments \cite{Giunti:2010zu},
which observed a deficit of electron neutrinos at distances of order 1~m from the radioactive source.
The anomaly has a statistical significance slightly smaller than $3\sigma$,
see also the recent analysis \cite{Kostensalo:2019vmv}.

In 2011, the updated calculations of the electron antineutrino fluxes
from nuclear reactors \cite{Mueller:2011nm,Huber:2011wv}
lead to the discovery of a new anomaly,
coming from a smaller observed event rate in a number of existing neutrino experiments
at reactors \cite{Mention:2011rk}
with respect to the predicted one.
Also in this case the combination of the various experiments gives a $\sim3\sigma$ significance.

A common solution of the two anomalies could come from a suppression of the measured flux due
to a non-zero effective disappearance angle $\vartheta_{ee}$,
but errors in the calculation of the unoscillated fluxes could also be viable explanations.
If the theoretical estimates of the reactor antineutrino flux were wrong,
for instance, the reactor anomaly would also be wrong.
Similar arguments could apply the Gallium anomaly.

In order to better investigate if neutrino oscillations could be
the possible explanation of these anomalies
and to reconstruct the reactor antineutrino flux with more precision,
in the recent years several experiments at SBL started to measure the antineutrino flux
at different distances from the reactor cores.
The obtained fluxes can then be used to compute ratios,
that in principle only depend on neutrino oscillation effects
and not on distance-independent systematics,
related for example to the normalization or shape of the unoscillated flux:
for this reason, the experiments of this class provide \emph{model-independent} results
(i.e.\ the theoretical model for the unoscillated flux is nearly irrelevant when computing the fit).

The first experiment to provide results obtained with a ratio method was
NEOS \cite{Ko:2016owz}, in South Korea,
which however has only one fixed detector at $\sim25$~m from the reactor core and
uses the DayaBay flux observations at much bigger distances (the reactor composition is similar)
in order to obtain the ratio.
A second experiment of this kind is
DANSS \cite{Alekseev:2018efk}, in Russia,
which instead has a movable detector that can be moved to three
different positions, at distances between $\sim10.5$ and $12.5$~m
from the reactor core.
From the fluxes at these three positions, two ratios can be computed.

The combined results of NEOS and DANSS,
until few months ago,
were indicating a preference in favor of
of the existence of new oscillations described by
$\dmij{41}\simeq1.3~\mbox{eV}^2$ and
$\Uaj{e4}\simeq0.01$
over the standard three neutrinos case,
with a global significance of $\sim3.5\sigma$
\cite{Gariazzo:2018mwd} (see also \cite{Dentler:2017tkw}).
Considering also the new set of data by the DANSS experiment
shown for the first time in the EPS-HEP conference \cite{danilov_epshep19},
together with
NEOS \cite{Ko:2016owz},
PROSPECT \cite{Ashenfelter:2018iov}
and
the previous DANSS \cite{Alekseev:2018efk}
observations,
the global model-independent preference in favor of the light sterile neutrino
decreases to a $\sim2.5\sigma$ significance and the new best-fit point,
nearly degenerate with the one mentioned above,
corresponds to
$\dmij{41}\simeq 0.4~$eV$^2$ and
$\Uaj{e4}\simeq0.01$
\cite{sterile19}.

Finally, the Neutrino-4 experiment \cite{Serebrov:2018vdw}
also provided results that indicate a strong preference in favor of a light sterile neutrino,
with a mass defined by a large $\dmij{41}\simeq7$~eV$^2$.
The best-fit of Neutrino-4, however,
is in direct tension with the constraints obtained by PROSPECT \cite{Ashenfelter:2018iov}
with the first 33 days of observations.

\subsection{Muon (anti)neutrino disappearance}
Considering muon (anti)neutrino disappearance,
no anomaly was ever observed.
Therefore, current experiments only provide strong upper bounds,
mainly on the matrix element \Uaj{\mu4}.

The bounds come from two classes of experiments:
atmospheric neutrino oscillation measurements,
mainly driven by the
IceCube \cite{Aartsen:2017bap,TheIceCube:2016oqi} observations,
and probes using accelerator neutrinos, such as in the MINOS+ case \cite{Adamson:2017uda}.
Atmospheric data, thanks to the strong matter effects that influence neutrino oscillations through Earth,
can constrain \Uaj{\mu4} and to a minor extent also \Uaj{\tau4},
in particular thanks to the low-energy data by the DeepCore section of the IceCube detector \cite{Aartsen:2017bap}.
Bounds on \Uaj{\mu4} are however approximately one order of magnitude stronger than those on \Uaj{\tau4}.

MINOS+ data \cite{Adamson:2017uda}, obtained by means of
a near ($\sim500$~m from the source)
and a far ($\sim800$~km) detector,
currently provide the strongest bounds on \Uaj{\mu4} in a wide range of \dmij{41} values.
Due to the distance between the source and the near detector,
MINOS+ can use a far-to-near flux ratio to constrain the neutrino mixing in a model-independent way
only for $\dmij{41}\lesssim1$~eV$^2$:
for larger mass splittings, there may be active-sterile neutrino oscillations already in the near detector
and it is impossible to measure the unoscillated flux.
For this reason, in the latest analyses the MINOS+ collaboration decided to use
a full two-detectors fit instead of a ratio fit.
In the high \dmij{41} range, the bounds have a significant dependence on cross-section systematics.
We have checked that, below $\dmij{41}\lesssim10$~eV$^2$,
which is the most interesting region given the results of reactor experiments,
a far-to-near ratio analysis gives results very similar
to those obtained with the full two-detectors fit \cite{sterile19}.
The treatment of systematic uncertainties, therefore,
does not affect significantly the obtained bound in the range $1\lesssim\dmij{41}/\mbox{eV}^2\lesssim10$.
We have also verified that the bounds on \Uaj{\mu4}
do not vary significantly when the three-neutrino mixing parameters or
the other active-sterile mixing angles are varied
in the analysis \cite{sterile19}.

\section{Appearance constraints}
The most controversial anomalies in SBL oscillations until now have been obtained
by (anti)neutrino appearance experiments, such as LSND \cite{Aguilar:2001ty}
and
MiniBooNE \cite{Aguilar-Arevalo:2018gpe}.

The LSND experiment was the first one to report the anomalous appearance of electron antineutrinos
in a beam of muon antineutrinos,
with a significance of $\sim3.8\sigma$.
Such anomaly was not confirmed by the KARMEN experiment,
working at slightly smaller distances \cite{Armbruster:2002mp}.

In order to test the LSND anomaly, the MiniBooNE experiment was built.
MiniBooNE uses neutrinos at higher energies with respect to LSND,
but it preserves the same $L/E$ of the anomaly.
The most recent MiniBooNE results \cite{Aguilar-Arevalo:2018gpe}
are in partial agreement with the LSND ones.
The preferred best-fit by MiniBooNE, however,
corresponds to maximal mixing between active and sterile states,
and is in direct tension with the
ICARUS \cite{Antonello:2013gut}
and
OPERA \cite{Agafonova:2013xsk}
results.
Moreover, even maximal mixing is not really sufficient to fully explain the excess
in the two bins at the lowest studied energies.
For this reason, a new experiment, MicroBooNE \cite{Chen:2007ae},
was proposed to check the LSND and MiniBooNE excess,
using liquid Argon time projection chamber (LArTPC) technology in order to be able to achieve
a better level of signal/background separation
and therefore understand if the anomalous events are really due to neutrino oscillations
or to some other kind of new physics.
MicroBooNE is also one of the three facilities that will constitute the SBL program at FermiLAB:
it will be the intermediate detector of the SBN experiment \cite{Antonello:2015lea}.

\section{Global fit}
As already anticipated in the introduction,
the three effective mixing angles which are mostly relevant for
electron (anti)neutrino disappearance ($\vartheta_{ee}$),
muon (anti)neutrino disappearance ($\vartheta_{\mu\mu}$)
and 
electron (anti)neutrino appearance ($\vartheta_{e\mu}$)
can be written in terms of two elements of the fourth column of the mixing matrix:
\Uaj{e4} and \Uaj{\mu4}.
In the ideal case, appearance and disappearance data
would indicate a common preferred region for such matrix elements and
we would have a single explanation for all the observed anomalies.
Unfortunately, this is not the case.

From the model-independent fit of NEOS and DANSS data we obtain
$\Uaj{e4}\simeq10^{-2}$, with a $3\sigma$ upper limit of about
$\Uaj{e4}\lesssim3\ex{-2}$.
From the muon disappearance channel,
mainly driven by MINOS+ and IceCube,
we have a $3\sigma$ upper bound of
$\Uaj{\mu4}\lesssim10^{-2}$
on the second entry of the last column of the mixing matrix.
Combining these two bounds from disappearance probes,
we expect
$\sin^22\vartheta_{e\mu}\lesssim10^{-3}$ at $3\sigma$.
In order to explain the anomaly observed by LSND and MiniBooNE,
on the other hand,
we would need a mixing angle
$\sin^22\vartheta_{e\mu}\gtrsim10^{-3}$, again at $3\sigma$.
Although these are approximate numbers,
they are sufficient to see that there is a tension between appearance and disappearance observations.

In order to quantify the tension between the two sets of constraints,
the easiest way is to adopt a parameter goodness of fit (PG) test
on the best-fit point.
The $p$-value of the PG for the full combination of appearance and disappearance data,
taking into account in particular the most recent results from MINOS+ and MiniBooNE,
is around $10^{-9}$ \cite{sterile19}, certainly too small to be due to random realizations
of the same underlying model.
We must conclude that nowadays there is no common sterile neutrino solution for the SBL anomalies
and some additional explanation is required in order to reconcile appearance and disappearance probes.

Using the PG, we can also test which experiment is mostly responsible for the tension \cite{sterile19}
(see also \cite{Dentler:2018sju}).
Since the muon disappearance experiments observe no anomaly,
and assuming that the model-independent observations of NEOS and DANSS
are not influenced by unaccounted systematics or new physics%
\footnote{Since their constraints are computed using ratio of spectra at different distances,
it is unlikely that some problem in the evaluation of the initial flux
or some new interactions can produce a distance-dependent effect
that simulates neutrino oscillations.},
we considered the effect of removing LSND, MiniBooNE or both of them from the analysis.
When the global fit is performed excluding MiniBooNE,
which claims to have a preference of $4.8\sigma$ in favor of the sterile neutrino presence,
the $p$-value becomes of the order of $10^{-6}$:
a significant improvement, but not sufficient to claim that the remaining appearance measurements
are compatible with disappearance probes.
On the other hand, we can remove LSND,
which reports a global preference of $3.8\sigma$ in favor of the 3+1 neutrinos case,
and in this case the $p$-value becomes approximately $10^{-5}$:
nearly an order of magnitude larger than in the case without MiniBooNE.
These numbers teach us that the preference for the 3+1 model from each experiment alone
does not reflect their role in the global fit:
LSND has a bigger effect on the global analysis because its best-fit
is not as much in tension with other experiments
as the MiniBooNE best-fit.
It is therefore inaccurate to claim that
MiniBooNE currently gives the strongest preference in favor of the new neutrino:
this is true only if all the other data are ignored.

The last test consists in removing both LSND and MiniBooNE from the global analysis.
In this case, there is no anomalous signal in the appearance channel,
so that the tension vanishes and the remaining experiments give a consistent fit
where \Uaj{\mu4} is compatible with zero and \Uaj{e4} is given by reactor experiments.
The fit obtained in this way can be motivated
by the possible existence of new physics beyond the light sterile neutrino:
if the LSND and MiniBooNE anomalies are not entirely due to active-sterile neutrino oscillations,
it is incorrect to include their data in a global fit of the 3+1 mixing parameters.
As already mentioned,
the MicroBooNE experiment and the SBN program
are expected to provide
a conclusive result on the subject in the next years.

\section{Light sterile neutrino and early Universe}
In addition to the tensions in neutrino oscillations,
another problem arises when a light sterile neutrino is considered.
The problem is related to the fact that,
if it exists,
a light sterile neutrino affects the evolution of the Universe
and cosmological observations can be used to put bounds on its properties.
In order to obtain these bounds,
one has to compute the effects of active-sterile neutrino oscillations in the early Universe
and determine if the sterile neutrino can reach equilibrium with the active flavors.
The thermalization process must be described in an environment
which contains the thermal plasma,
composed by muons, electrons, photons and neutrinos.
The calculation must take into account the expansion of the Universe,
annihilation processes which transfer energy from muons and electrons to the rest of the thermal plasma,
energy transfer between neutrinos and electrons
and of course neutrino oscillations.

When the various particles are in equilibrium,
they have a Fermi-Dirac or Bose-Einstein distribution function,
but the crucial point is that the neutrino momentum distribution
is not necessarily the equilibrium one, for two reasons.
First, the sterile neutrino is not expected to exist in the very early Universe,
because it cannot be generated by the electroweak processes that keep the plasma in equilibrium:
it must be produced by oscillations once the matter effects become small enough
to allow active-sterile neutrino oscillations.
This means that the distribution function of the fourth neutrino
is initially zero and evolves in a non trivial way
towards its final shape, which can be different from a pure Fermi-Dirac.
Second, after most of the neutrinos have decoupled from the thermal plasma,
electrons annihilate and transfer energy to the photons and to the few neutrinos still coupled to them,
those in the high-momentum tail of the momentum distribution,
which is distorted by the energy transfer.
In order to compute these effects,
one possible approach is to discretise the neutrino distribution function
using a grid of momenta, and evolve its value in each point of such grid
independently \cite{Gariazzo:2019gyi}.

In practice,
in order to obtain the final momentum distribution function of the various particles in the thermal plasma
one has to solve a differential equation that governs the evolution of the photon temperature
plus a set of equations which describe the evolution of the neutrino momentum distribution
for the various flavors, as a function of the neutrino momentum~\cite{Gariazzo:2019gyi}.
Solving these equations
taking into account the preferred mass splitting and mixing angles
that emerge from SBL observations,
for example the values obtained combining DANSS and NEOS results,
one obtains that the additional neutrino reaches a full thermalization
before the interactions between neutrinos and electrons become weak enough
and the neutrinos decouple from the thermal plasma.
In terms of the widely used effective number of relativistic species, \Neff,
all the points within the preferred region at $3\sigma$ from the DANSS+NEOS experiments
correspond to $\Neff\simeq4$~\cite{Gariazzo:2019gyi}.

Cosmological bounds, however, prefer a much smaller \Neff.
For instance, Big Bang Nucleosynthesis data constrain
$\Neff\simeq2.9\pm0.2$ \cite{Peimbert:2016bdg},
while from Cosmic Microwave Background (CMB) observations one obtains
$\Neff\lesssim3.3$ \cite{Aghanim:2018eyx}.
Going more in details,
CMB data are compatible with a sterile-neutrino-like particle
with a mass around 1~eV only if its contribution to \Neff\ is very small,
or with a somewhat larger \Neff\ only if it comes from nearly massless particles
(see e.g.\ \cite{Aghanim:2018eyx,Archidiacono:2016kkh}).

If the future reactor neutrino experiments will confirm the current best-fit point,
therefore, some new physics will be required in order to reconcile the presence of the light sterile neutrino
in the early Universe with the current observations.
Such new physics may be in the form of new interactions (see e.g.\ \cite{Archidiacono:2016kkh}),
preventing the thermalization of the new neutrino thanks to the presence of additional matter effects,
which suppress neutrino oscillations between the active and the sterile states at the relevant times.

\section{Prospects and conclusions}
In the incoming months, many experiments are expected to publish more results.
In particular,
apart for the DANSS experiment that is still taking data \cite{danilov_epshep19},
results are expected from
STEREO \cite{Almazan:2018wln,Bernard:2019jli}
and
PROSPECT \cite{Ashenfelter:2018iov}.
Currently, the limits from STEREO and PROSPECT are not competitive enough to confirm or reject
the preferred oscillation parameters by DANSS and NEOS,
but they are expected to reach soon the required sensitivity.
If within the next few years the different experiment will not converge towards a common best-fit point,
the light sterile neutrino explanation of the anomalies will need to be discarded.
On the other hand,
if many of them will independently observe oscillations involving a new neutrino state,
with the same mixing parameters,
we will have the cleanest signal ever observed in favor of new physics beyond the standard model.

In the same way, the already mentioned
MicroBooNE \cite{Chen:2007ae}
and
SBN \cite{Antonello:2015lea}
experiments
will soon be able to give a final confirmation or disproval
of the sterile neutrino interpretation of the LSND and MiniBooNE results.

A confirmation of the light sterile neutrino from oscillation experiments,
moreover, will require some new mechanism in order to reconcile the presence of the new particle
in the early Universe with the current observational bounds.
All together, these new experiments will therefore allow us to understand whether
a consistent explanation of the SBL anomalies can exist or not,
and, if it may involve a light sterile neutrino,
which are the mixing parameters associated to it,
shedding light on what stands beyond the standard model of particle physics.

\section*{Acknowledgements}
The author receives support from the European Union's Horizon 2020 research and innovation programme under the Marie Sk{\l}odowska-Curie individual grant agreement No.\ 796941.

\bibliographystyle{JHEP}
\bibliography{main}

\providecommand{\href}[2]{#2}\begingroup\raggedright\begin{thebibliography}{10}

\bibitem{deSalas:2017kay}
P.~de~Salas, D.~Forero, C.~Ternes, M.~T{\'o}rtola and J.~Valle, \emph{{Status
  of neutrino oscillations 2018: 3 $\sigma$ hint for normal mass ordering and
  improved {CP} sensitivity}},
  \href{https://doi.org/10.1016/j.physletb.2018.06.019}{\emph{Phys.Lett.}
  {\bfseries B782} (2018) 633}
  [\href{https://arxiv.org/abs/1708.01186}{{\ttfamily 1708.01186}}].

\bibitem{Aguilar:2001ty}
{\scshape LSND} collaboration, A.~Aguilar-Arevalo et~al., \emph{{Evidence for
  neutrino oscillations from the observation of anti-neutrino(electron)
  appearance in a anti-neutrino(muon) beam}},
  \href{https://doi.org/10.1103/PhysRevD.64.112007}{\emph{Phys.Rev.D}
  {\bfseries 64} (2001) 112007}
  [\href{https://arxiv.org/abs/hep-ex/0104049}{{\ttfamily hep-ex/0104049}}].

\bibitem{Giunti:2010zu}
C.~Giunti and M.~Laveder, \emph{{Statistical Significance of the Gallium
  Anomaly}},
  \href{https://doi.org/10.1103/PhysRevC.83.065504}{\emph{Phys.Rev.C}
  {\bfseries 83} (2011) 065504}
  [\href{https://arxiv.org/abs/1006.3244}{{\ttfamily 1006.3244}}].

\bibitem{Kostensalo:2019vmv}
J.~Kostensalo, J.~Suhonen, C.~Giunti and P.~C. Srivastava, \emph{{The gallium
  anomaly revisited}},
  \href{https://doi.org/10.1016/j.physletb.2019.06.057}{\emph{Phys.Lett.B}
  {\bfseries 795} (2019) 542}
  [\href{https://arxiv.org/abs/1906.10980}{{\ttfamily 1906.10980}}].

\bibitem{Mention:2011rk}
G.~Mention, M.~Fechner, T.~Lasserre, T.~A. Mueller, D.~Lhuillier, M.~Cribier
  et~al., \emph{{The Reactor Antineutrino Anomaly}},
  \href{https://doi.org/10.1103/PhysRevD.83.073006}{\emph{Phys.Rev.D}
  {\bfseries 83} (2011) 073006}
  [\href{https://arxiv.org/abs/1101.2755}{{\ttfamily 1101.2755}}].

\bibitem{Mueller:2011nm}
T.~A. Mueller et~al., \emph{{Improved Predictions of Reactor Antineutrino
  Spectra}},
  \href{https://doi.org/10.1103/PhysRevC.83.054615}{\emph{Phys.Rev.C}
  {\bfseries 83} (2011) 054615}
  [\href{https://arxiv.org/abs/1101.2663}{{\ttfamily 1101.2663}}].

\bibitem{Huber:2011wv}
P.~Huber, \emph{{On the determination of anti-neutrino spectra from nuclear
  reactors}},
  \href{https://doi.org/10.1103/PhysRevC.84.024617}{\emph{Phys.Rev.C}
  {\bfseries 84} (2011) 024617}
  [\href{https://arxiv.org/abs/1106.0687}{{\ttfamily 1106.0687}}].

\bibitem{ALEPH:2005ab}
{\scshape SLD Electroweak Group, DELPHI, ALEPH, SLD, SLD Heavy Flavour Group,
  OPAL, LEP Electroweak Working Group, L3} collaboration, S.~Schael et~al.,
  \emph{{Precision electroweak measurements on the $Z$ resonance}},
  \href{https://doi.org/10.1016/j.physrep.2005.12.006}{\emph{Phys.Rept.}
  {\bfseries 427} (2006) 257}
  [\href{https://arxiv.org/abs/hep-ex/0509008}{{\ttfamily hep-ex/0509008}}].

\bibitem{Gariazzo:2015rra}
S.~Gariazzo, C.~Giunti, M.~Laveder, Y.~F. Li and E.~M. Zavanin, \emph{{Light
  sterile neutrinos}},
  \href{https://doi.org/10.1088/0954-3899/43/3/033001}{\emph{J.Phys.G}
  {\bfseries 43} (2016) 033001}
  [\href{https://arxiv.org/abs/1507.08204}{{\ttfamily 1507.08204}}].

\bibitem{Ko:2016owz}
Y.~J. Ko et~al., \emph{{Sterile Neutrino Search at the NEOS Experiment}},
  \href{https://doi.org/10.1103/PhysRevLett.118.121802}{\emph{Phys.Rev.Lett.}
  {\bfseries 118} (2017) 121802}
  [\href{https://arxiv.org/abs/1610.05134}{{\ttfamily 1610.05134}}].

\bibitem{Alekseev:2018efk}
{\scshape DANSS} collaboration, I.~Alekseev et~al., \emph{{Search for sterile
  neutrinos at the DANSS experiment}},
  \href{https://doi.org/10.1016/j.physletb.2018.10.038}{\emph{Phys.Lett.}
  {\bfseries B787} (2018) 56}
  [\href{https://arxiv.org/abs/1804.04046}{{\ttfamily 1804.04046}}].

\bibitem{Gariazzo:2018mwd}
S.~Gariazzo, C.~Giunti, M.~Laveder and Y.~F. Li, \emph{{Model-Independent
  $\bar\nu_{e}$ Short-Baseline Oscillations from Reactor Spectral Ratios}},
  \href{https://doi.org/10.1016/j.physletb.2018.04.057}{\emph{Phys.Lett.}
  {\bfseries B782} (2018) 13}
  [\href{https://arxiv.org/abs/1801.06467}{{\ttfamily 1801.06467}}].

\bibitem{Dentler:2017tkw}
M.~Dentler, A.~Hern\'{a}ndez-Cabezudo, J.~Kopp, M.~Maltoni and T.~Schwetz,
  \emph{{Sterile Neutrinos or Flux Uncertainties? - Status of the Reactor
  Anti-Neutrino Anomaly}},
  \href{https://doi.org/10.1007/JHEP11(2017)099}{\emph{JHEP} {\bfseries 1711}
  (2017) 099} [\href{https://arxiv.org/abs/1709.04294}{{\ttfamily
  1709.04294}}].

\bibitem{danilov_epshep19}
M.~Danilov, ``{New results from the DANSS experiment}.'' Talk at \emph{EPS-HEP
  2019}, 10--17 July 2019, Ghent, Belgium.

\bibitem{Ashenfelter:2018iov}
{\scshape PROSPECT} collaboration, J.~Ashenfelter et~al., \emph{{First search
  for short-baseline neutrino oscillations at HFIR with PROSPECT}},
  \href{https://doi.org/10.1103/PhysRevLett.121.251802}{\emph{Phys.Rev.Lett.}
  {\bfseries 121} (2018) 251802}
  [\href{https://arxiv.org/abs/1806.02784}{{\ttfamily 1806.02784}}].

\bibitem{sterile19}
S.~Gariazzo, C.~Giunti and C.~Ternes. in preparation.

\bibitem{Serebrov:2018vdw}
{\scshape NEUTRINO-4} collaboration, A.~Serebrov et~al., \emph{{The first
  observation of effect of oscillation in Neutrino-4 experiment on search for
  sterile neutrino}},
  \href{https://doi.org/10.1134/S0021364019040040}{\emph{Pisma
  Zh.Eksp.Teor.Fiz.} {\bfseries 109} (2019) 209}
  [\href{https://arxiv.org/abs/1809.10561}{{\ttfamily 1809.10561}}].

\bibitem{Aartsen:2017bap}
{\scshape IceCube} collaboration, M.~G. Aartsen et~al., \emph{{Search for
  sterile neutrino mixing using three years of IceCube DeepCore data}},
  \href{https://doi.org/10.1103/PhysRevD.95.112002}{\emph{Phys.Rev.D}
  {\bfseries 95} (2017) 112002}
  [\href{https://arxiv.org/abs/1702.05160}{{\ttfamily 1702.05160}}].

\bibitem{TheIceCube:2016oqi}
{\scshape IceCube} collaboration, M.~G. Aartsen et~al., \emph{{Searches for
  Sterile Neutrinos with the IceCube Detector}},
  \href{https://doi.org/10.1103/PhysRevLett.117.071801}{\emph{Phys.Rev.Lett.}
  {\bfseries 117} (2016) 071801}
  [\href{https://arxiv.org/abs/1605.01990}{{\ttfamily 1605.01990}}].

\bibitem{Adamson:2017uda}
{\scshape MINOS} collaboration, P.~Adamson et~al., \emph{{Search for sterile
  neutrinos in MINOS and MINOS+ using a two-detector fit}},
  \href{https://doi.org/10.1103/PhysRevLett.122.091803}{\emph{Phys.Rev.Lett.}
  {\bfseries 122} (2019) 091803}
  [\href{https://arxiv.org/abs/1710.06488}{{\ttfamily 1710.06488}}].

\bibitem{Aguilar-Arevalo:2018gpe}
{\scshape MiniBooNE} collaboration, A.~Aguilar-Arevalo et~al.,
  \emph{{Observation of a Significant Excess of Electron-Like Events in the
  MiniBooNE Short-Baseline Neutrino Experiment}},
  \href{https://doi.org/10.1103/PhysRevLett.121.221801}{\emph{Phys.Rev.Lett.}
  {\bfseries 121} (2018) 221801}
  [\href{https://arxiv.org/abs/1805.12028}{{\ttfamily 1805.12028}}].

\bibitem{Armbruster:2002mp}
{\scshape KARMEN} collaboration, B.~Armbruster et~al., \emph{{Upper limits for
  neutrino oscillations muon-anti-neutrino $\rightarrow$ electron-anti-neutrino
  from muon decay at rest}},
  \href{https://doi.org/10.1103/PhysRevD.65.112001}{\emph{Phys.Rev.} {\bfseries
  D65} (2002) 112001} [\href{https://arxiv.org/abs/hep-ex/0203021}{{\ttfamily
  hep-ex/0203021}}].

\bibitem{Antonello:2013gut}
{\scshape ICARUS} collaboration, M.~Antonello et~al., \emph{{Search for
  anomalies in the ${\nu}_e$ appearance from a ${\nu}_{\mu}$ beam}},
  \href{https://doi.org/10.1140/epjc/s10052-013-2599-z}{\emph{Eur.Phys.J.C}
  {\bfseries 73} (2013) 2599}
  [\href{https://arxiv.org/abs/1307.4699}{{\ttfamily 1307.4699}}].

\bibitem{Agafonova:2013xsk}
{\scshape OPERA} collaboration, N.~Agafonova et~al., \emph{{Search for $\nu_\mu
  \rightarrow \nu_e$ oscillations with the OPERA experiment in the CNGS beam}},
  \href{https://doi.org/10.1007/JHEP07(2013)085}{\emph{JHEP} {\bfseries 07}
  (2013) 004} [\href{https://arxiv.org/abs/1303.3953}{{\ttfamily 1303.3953}}].

\bibitem{Chen:2007ae}
{\scshape MicroBooNE} collaboration, H.~Chen et~al., \emph{{Proposal for a New
  Experiment Using the Booster and NuMI Neutrino Beamlines: MicroBooNE}}, .

\bibitem{Antonello:2015lea}
{\scshape MicroBooNE} collaboration, M.~Antonello et~al., \emph{{A Proposal for
  a Three Detector Short-Baseline Neutrino Oscillation Program in the Fermilab
  Booster Neutrino Beam}},  \href{https://arxiv.org/abs/1503.01520}{{\ttfamily
  1503.01520}}.

\bibitem{Dentler:2018sju}
M.~Dentler et~al., \emph{{Updated Global Analysis of Neutrino Oscillations in
  the Presence of eV-Scale Sterile Neutrinos}},
  \href{https://doi.org/10.1007/JHEP08(2018)010}{\emph{JHEP} {\bfseries 08}
  (2018) 010} [\href{https://arxiv.org/abs/1803.10661}{{\ttfamily
  1803.10661}}].

\bibitem{Gariazzo:2019gyi}
S.~Gariazzo, P.~de~Salas and S.~Pastor, \emph{{Thermalisation of sterile
  neutrinos in the early Universe in the 3+1 scheme with full mixing matrix}},
  \href{https://doi.org/10.1088/1475-7516/2019/07/014}{\emph{JCAP} {\bfseries
  1907} (2019) 014} [\href{https://arxiv.org/abs/1905.11290}{{\ttfamily
  1905.11290}}].

\bibitem{Peimbert:2016bdg}
A.~Peimbert, M.~Peimbert and V.~Luridiana, \emph{{The primordial helium
  abundance and the number of neutrino families}},
  {\emph{Rev.Mex.Astron.Astrofis.} {\bfseries 52} (2016) 419}
  [\href{https://arxiv.org/abs/1608.02062}{{\ttfamily 1608.02062}}].

\bibitem{Aghanim:2018eyx}
{\scshape Planck} collaboration, N.~Aghanim et~al., \emph{{Planck 2018 results.
  VI. Cosmological parameters}},
  \href{https://arxiv.org/abs/1807.06209}{{\ttfamily 1807.06209}}.

\bibitem{Archidiacono:2016kkh}
M.~Archidiacono, S.~Gariazzo, C.~Giunti, S.~Hannestad, R.~Hansen, M.~Laveder
  et~al., \emph{{Pseudoscalar--sterile neutrino interactions: reconciling the
  cosmos with neutrino oscillations}},
  \href{https://doi.org/10.1088/1475-7516/2016/08/067}{\emph{JCAP} {\bfseries
  08} (2016) 067} [\href{https://arxiv.org/abs/1606.07673}{{\ttfamily
  1606.07673}}].

\bibitem{Almazan:2018wln}
{\scshape STEREO} collaboration, H.~Almaz{\'a}n et~al., \emph{{Sterile Neutrino
  Constraints from the STEREO Experiment with 66~Days of Reactor-On Data}},
  \href{https://doi.org/10.1103/PhysRevLett.121.161801}{\emph{Phys.Rev.Lett.}
  {\bfseries 121} (2018) 161801}
  [\href{https://arxiv.org/abs/1806.02096}{{\ttfamily 1806.02096}}].

\bibitem{Bernard:2019jli}
{\scshape STEREO} collaboration, L.~Bernard, \emph{{Results from the STEREO
  Experiment with 119 days of Reactor-on Data}},  2019,
  \href{https://arxiv.org/abs/1905.11896}{{\ttfamily 1905.11896}}.

\end{thebibliography}\endgroup

\end{document}